\documentclass[prd,nofootinbib,showpacs,twocolumn]{revtex4-1}

\usepackage{graphicx}
\usepackage{float}
\usepackage{amsmath}
\usepackage{cancel}
\usepackage{hyperref}
\usepackage{mathtools}
\usepackage{txfonts}

\newcommand{\diffd}{\textrm{d}}

\begin{document}

\title{Phenomenological model for the Drell-Yan process: Reexamined}
\author{Fabian Eichstaedt}\email{fabian.eichstaedt@theo.physik.uni-giessen.de}
\author{Stefan Leupold}
\author{Ulrich Mosel}
\affiliation{Institut f\"ur Theoretische Physik, Universit\"at Giessen, Germany}

\begin{abstract}
Drell-Yan pair production is investigated. We reexamine a model where the quark momentum fraction 
is defined as the ratio of the corresponding light cone components of the quark and parent nucleon 
in a naive parton-model approach. It is shown that results differ from the
standard parton model. This is due to unphysical solutions for the momentum
fractions within the naive approach which are not present in the standard parton model.
In a calculation employing full quark kinematics, i.e. including primordial quark transverse momentum,
these solutions also appear. A prescription is given to handle these solutions in order to avoid incorrect
results. The impact of these solutions in the full kinematical approach is demonstrated
and compared to the modified result.

\end{abstract}



\maketitle

\section{Introduction}
\label{sec:intro}
The Drell-Yan (DY) process \cite{Drell:1970wh} was first described in the 1970s and provides
an important tool to access the distribution of partons inside the nucleon.
While a lot of information can be gained from deep inelastic scattering \cite{Bjorken:1969ja},
measurements of Drell-Yan events give complementary insights, especially about sea-quark
distributions \cite{Alekhin:2006zm}. This has sparked many studies of this process 
\cite{Halzen:1978rx,Altarelli:1977kt,Arnold:2008kf}
which are generally inspired by perturbative QCD (pQCD). A lot of experimental effort
is being devoted to measurements of the DY process: In antiproton-proton collisions at
$\overline{\textrm{P}}$ANDA (FAIR) \cite{Lutz:2009ff} and PAX \cite{Barone:2005pu}, in proton-proton
collisions at RHIC \cite{Bunce:2000uv,Bunce:2008aa}, J-PARC 
\cite{Peng:2006aa,Goto:2007aa,Kumano:2008rt}, IHEP \cite{Abramov:2005mk} and JINR
\cite{Sissakian:2008th} and in pion-nucleon collisions at COMPASS \cite{Bradamante:1997mu,Baum:1996yv}.
An overview of the experimental situation can be found in \cite{Reimer:2007iy}.
$\overline{\textrm{P}}$ANDA, for example, will allow measurements at hadron c.m. energies
of a few GeV, where non-perturbative effects are expected to become more important.
This highlights the need to model these effects in a phenomenological picture.

In addition the standard pQCD leading order (i.e. parton model) description 
does not fully describe the interesting observables. Invariant
mass ($M$) spectra of the DY pair can only be accounted for by including an additional 
K-factor and transverse momentum ($p_T$) spectra are not accessible at all 
\cite{Gavin:1995ch}.
The latter can be
partly cured by folding in a phenomenological Gaussian distribution for the transverse
momentum, the width of which has to be fitted to data. However the absolute size of the
cross sections is still underestimated \cite{D'Alesio:2004up}. Next-to-leading-order (NLO)
calculations improve the description in some aspects, but also bring about additional problems.
The calculated invariant mass spectra come closer to the data and the $p_T$ spectra are
comparable to data in the region $p_T \sim M$ \cite{Gavin:1995ch}, but not for 
$p_T \rightarrow 0$ \cite{Halzen:1978rx}. In fact the $p_T$-spectrum is divergent for 
$p_T \rightarrow 0$ in any fixed order of the strong coupling $\alpha_s$,
due to large logarithmic corrections $\ln \left(M/p_T\right)$. These stem from soft gluon exchange
and it is possible to remove these divergencies by an all-order resummation. However since
$p_T$ is no longer a hard scale at $p_T \rightarrow 0$ additional non-pertubative (i.e.
experimental) input 
is needed in these (and all other pQCD) approaches to describe the region of very small $p_T$ 
\cite{Collins:1984kg,Davies:1984sp,Fai:2003zc}. Note here that the parton model (i.e.
leading order) description is still a very useful starting point, e.g. for studying 
spin asymmetries in DY, since there NLO corrections appear to be rather small 
\cite{Shimizu:2005fp,Barone:2005cr,Martin:1997rz}.

A phenomenological model that incorporates full transverse momentum dependent quark 
kinematics and which in addition allows for mass distributions of quarks was proposed to 
resolve these problems \cite{Linnyk:2004mt,Linnyk:2005iw,Linnyk:2006mv}. The idea stems
from the fact that in the usual collinear approach the parton momenta are confined to
the beam direction, thus only one momentum component is different from zero. The other
components, namely the transverse momentum and the mass of the parton, do not enter
into the calculation of the partonic subprocess cross section. Since at finite energies
these components might influence the cross section to some extent it is worthwhile to examine
this influence in detail. However it turns out that in these works
important physical constraints were not considered and thus incorrect results were obtained.
In the current paper we examine
these constraints in detail and present a prescription to properly account for them.
Finally we compare the results of the treatment in 
\cite{Linnyk:2004mt,Linnyk:2005iw,Linnyk:2006mv} with our corrected results. Since the
mentioned problems already appear for the case without mass distributions for partons we
restrict ourselves here to massless partons.

This paper is organized as follows: in Sec. \ref{sec:col} we compare the standard
collinear parton model description for DY with an approach that defines the parton momentum
fraction $x$ via light cone components. The latter approach will be a demonstration
of the problems that appear in the calculation with the full kinematics. Section \ref{sec:full}
contains two calculations in the full kinematical scheme, i.e.\ taking into account the full
transverse momentum dependence of the partonic subprocess. The approach of 
\cite{Linnyk:2004mt,Linnyk:2005iw,Linnyk:2006mv}
is discussed in detail in Sec. \ref{subsubsec:naive} and the technical details are given
in the Appendix.  
Section \ref{subsubsec:correct} then contains our calculation
which respects the physical constraints laid out in Sec. \ref{subsec:npm}.
The numerical results are presented in Sec. \ref{sec:res} where we compare
the two calculations of Sec. \ref{sec:full} quantitatively. Finally we present our
conclusions in Sec. \ref{sec:conc}.

\begin{figure}[htbp]
	\centering
	\includegraphics[keepaspectratio,width=0.45\textwidth]{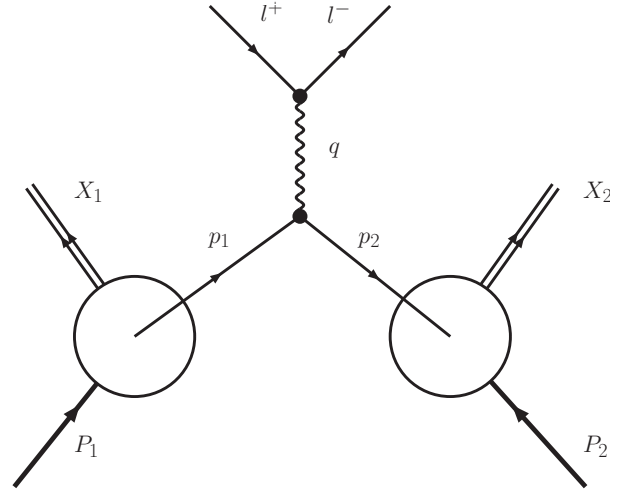}
	\caption{DY production in a nucleon-nucleon collision; $X_1$ and $X_2$ denote the
		 nucleon remnants. See main text for details.}
	\label{fig:DY}	 
\end{figure}

In the following we present the conventions and notations used throughout
this paper:
It will turn out to be useful to write four-momenta using light-cone coordinates. 
We employ the following convention for general four-vectors $a$ and $b$

\begin{align}
	           a^+ &= a_0 + a_z\ , \\
	           a^- &= a_0 - a_z\ , \\
	\vec a_{\perp} &= \left( a_x, a_y \right)\ , \\
       \Rightarrow a^2 &= a^+ a^- - ({\vec a}_{\perp})^2\ , \\
  \Rightarrow a\cdot b &= \frac{1}{2}\left(a^+ b^- + a^- b^+ - 2\, \vec a_{\perp} \cdot 
		                                                   \vec b_{\perp} \right)\ .
\end{align}       	
We regard all particles as massless. We define the target nucleon to carry the four-momentum 
$P_1$ and the beam nucleon to carry the four-momentum $P_2$ (see fig.\ \ref{fig:DY}). In the hadron
center-of-mass (c.m.) frame we choose the $z$-axis as the beam line and the beam (target) nucleon
moves in the positive (negative) direction.
Therefore the nucleon four-momenta read
\begin{align}
	P_1 &= \left( \frac{\sqrt{S}}{2} , 0, 0, -\frac{\sqrt{S}}{2} \right)\ , \label{nucleon1}\\
	P_2 &= \left( \frac{\sqrt{S}}{2} , 0, 0, +\frac{\sqrt{S}}{2} \right)\ . \label{nucleon2}
\end{align}
Note here that with a finite nucleon mass $P_1$ and $P_2$ would change. We have explicitly
conducted the entire calculation with non-zero nucleon mass and convinced ourselves that
it does not influence our arguments in Secs. \ref{sec:col} and \ref{sec:full}. Our
results in Sec. \ref{sec:res} would receive only a small correction, since we are looking
at c.m. energies of $\sqrt{S} > 27$ GeV. Thus, in this paper, we have put the nucleon mass
to zero for the sake of simplicity and readability.

We denote the four-momentum of the parton in nucleon 1 (2) as $p_1$ ($p_2$). The on-shell
condition in light-cone coordinates then reads
\begin{equation}
	0 = p_i^2 = p_i^+ p_i^- - ({\vec p}_{i_\perp})^2\ .
\end{equation}	
The definition of the Feynman variable $x_F$ is \cite{Amsler:2008zzb}
\begin{equation}
	x_F = \frac{q_z}{(q_z)_\textrm{max}} \ . \label{xF}
\end{equation}
For the virtual photon in fig.\ \ref{fig:DY} the maximal $q_z$ is derived by requiring the invariant
mass of the undetected remnants to vanish and the photon to move collinearly to the nucleons:
\begin{align}
	\left(P_1 + P_2 - q \right)^2 &= X^2 \overset{!}{=} 0 \\
	\Rightarrow S + q^2 - 2 \sqrt{S} \sqrt{q^2 + (q_z)_\textrm{max}^2} &= 0 \\
	\Rightarrow \frac{S - q^2}{2\sqrt{S}} &= (q_z)_\textrm{max} \label{qzmax} \ .
\end{align}

\section{Collinear approach}
\label{sec:col}

In this section we treat the interacting partons as collinear with their parent nucleons. We
compare the standard textbook parton model with a naive approach which uses the
light-cone component definition of the parton momentum fractions. It will turn
out that in the latter case unphysical solutions appear that must be removed to be 
consistent with the standard parton model.

\subsection{Standard parton model}
\label{subsec:spm}
The leading-order Drell-Yan total differential cross section in the standard parton
model reads \cite{Drell:1970wh}

\begin{equation}
	\textrm{d}\sigma = \int_0^1 \textrm{d}x_1 \int_0^1 \textrm{d}x_2
		                  \sum_i q_i^2\, f_i(x_1,q^2)\, f_{\bar i}(x_2,q^2)\, 
				  \textrm{d}\hat\sigma(x_1,x_2,q^2)\ .  \label{hadr_coll_xsec}
\end{equation}
Here $x_1$ and $x_2$ are the momentum fractions carried by the
annihilating partons inside the colliding nucleons:
\begin{align}
	p_1 &= x_1 P_1\ , \label{collinear_p1}  \\
	p_2 &= x_2 P_2\ . \label{collinear_p2}
\end{align}	
The sum runs over all quark flavors and
antiflavors, $q_i$ denotes the electric charge of quark flavor $i$, the functions
$f_i$ are parton distribution functions (PDFs) and $\textrm{d}\hat\sigma$ is the total differential
cross section of the partonic subprocess,
\begin{equation}
	\textrm{d}\hat \sigma = \frac{4 \pi \alpha^2}{9 q^2} \, \delta(M^2 - q^2) \,
				\delta^{(4)}(p_1 + p_2 - q) \, \textrm{d}^4q \, \textrm{d}M^2 \ .
						\label{part_coll_xsec}
\end{equation}				
Here $q$ is the four-momentum of the virtual photon, $p_1,p_2$ are the four-momenta of the
partons (cf.\ fig.\ \ref{fig:DY}) and $\alpha \approx 1/137$ is the fine-structure constant.

Note that it becomes immediately clear from Eqs.\ (\ref{collinear_p1}) and (\ref{collinear_p2})
that the incoming partons move collinearly with the nucleons. According to Eq.\ 
(\ref{part_coll_xsec}) no transverse momentum can be
generated for the virtual photon (and thus for the DY pair) in the leading-order process:
\begin{align}
	{\vec p}_{1_\perp} &= {\vec p}_{2_\perp} = 0 \\
	\Rightarrow \delta^{(4)}(p_1 + p_2 - q) &= \delta ( (p_1 + p_2)^0 - q^0)\,
						  \delta^{(2)}( {\vec q}_\perp ) \nonumber \\
						&\, \times  \delta ( (p_1 + p_2)^z - q^z ) \ .
\end{align}

The maximal information about the DY pair that can be gained from Eq.\ 
(\ref{hadr_coll_xsec}) is double differential. A common choice of variables is the
squared invariant mass $M^2$ and Feynman's $x_F$ of the virtual photon:
\begin{widetext}
\begin{align}
	\frac{\textrm{d}\hat \sigma}{\diffd M^2\ \diffd x_F} &= 
				\int \diffd^4q
                                \frac{4 \pi \alpha^2}{9 q^2} \, \delta(M^2 - q^2) \,
				\delta^{(4)}(p_1 + p_2 - q) \, 
				\delta\left(x_F - \frac{q_z}{(q_z)_\textrm{max}}\right)
	                                                     = 
                                \frac{4 \pi \alpha^2}{9 M^2} \, 
				\delta\left(M^2 - (p_1+p_2)^2\right)\,
				\delta\left(x_F - \frac{(p_1 + p_2)_z}{(q_z)_\textrm{max}}\right)
				\ .
\end{align}
\end{widetext}
The two $\delta$-functions connect $x_1$ and $x_2$ with the chosen observables
\begin{align}
	M^2 &= 2 p_1 p_2 = x_1 x_2 S \ , \label{M2_pm} \\
	x_F &= + \frac{\sqrt{S}\, (x_2 - x_1) }{2 (q_z)_\textrm{max} } \label{xF_pm}
\end{align}
with $(q_z)_\textrm{max} = \frac{S - M^2}{2\sqrt{S}}$, cf.\ Eq.\ (\ref{qzmax}).

Solving for $x_1$ and $x_2$ yields
\begin{align}
	x_{1_\pm} &= \frac{- (q_z)_\textrm{max}\, x_F \pm 
	\sqrt{ \left((q_z)_\textrm{max}\, x_F\right)^2 + M^2 }}{\sqrt{S}} \nonumber \\
	&= \frac{- (q_z)_\textrm{max}\, x_F \pm E_\textrm{coll}}{\sqrt{S}} \ ,
\end{align}
\begin{align}
	x_{2_\pm} &= \frac{(q_z)_\textrm{max}\, x_F \pm 
	\sqrt{ \left((q_z)_\textrm{max}\, x_F\right)^2 + M^2 }}{\sqrt{S}} \nonumber \\
	&= \frac{(q_z)_\textrm{max}\, x_F \pm E_\textrm{coll}}{\sqrt{S}}\ ,
\end{align}	
with the energy of the collinear DY-pair
\begin{equation}
E_\textrm{coll} = \sqrt{M^2 + \left((q_z)_\textrm{max}\, x_F\right)^2}\ .
\end{equation}
However the lower solutions are always negative. Only the upper solutions are
in the integration range of Eq.\ (\ref{hadr_coll_xsec}) and are physically meaningful.
For the negative solutions the parton energies would be negative on account of 
Eqs.\ (\ref{collinear_p1}) and (\ref{collinear_p2}).
The hadronic cross section then reads:
\begin{widetext}
\begin{align}
\frac{\diffd\sigma}{\diffd M^2\ \diffd x_F}
         	&= \int_0^1 \textrm{d}x_1 \int_0^1 \textrm{d}x_2
		                  \sum_i q_i^2 f_i(x_1,M^2) \, f_{\bar i}(x_2,M^2) \,
				  \frac{4 \pi \alpha^2}{9 M^2} \, 
				  \delta\left(M^2 - (p_1+p_2)^2\right)\,
				  \delta\left(x_F - \frac{(p_1 + p_2)_z}{(q_z)_\textrm{max}}\right)
				  \nonumber \\
	        &= \int_0^1 \textrm{d}x_1 \int_0^1 \textrm{d}x_2
		                  \sum_i q_i^2 f_i(x_1,M^2) \, f_{\bar i}(x_2,M^2) \,
				  \frac{4 \pi \alpha^2}{9 M^2} \, 
				  \frac{(q_z)_\textrm{max}}
                                       {S\sqrt{(q_z)^2_\textrm{max} x_F^2 + M^2} }
				  \delta(x_1 - x_{1_+})\,
				  \delta(x_2 - x_{2_+})
				  \nonumber \\
		&= \sum_i q_i^2 f_i(x_{1_+},M^2) \, f_{\bar i}(x_{2_+},M^2) \,
				  \frac{4 \pi \alpha^2}{9 M^2} \, 
				  \frac{(q_z)_\textrm{max}}
                                       {S E_\textrm{coll} } \ .
		\label{spm_hadr_xsec}
\end{align}
\end{widetext}

In this section we have presented the standard parton model solution for the leading-order DY
cross section. The only quantities in this approach not determined by pQCD are the PDFs.
These have to be obtained by fitting parametrizations to experimental data, mainly on deep inelastic
scattering (DIS), but also on measurements of DY production itself \cite{Stirling:1900sj}.

\subsection{Naive parton model}
\label{subsec:npm}

In this section we work out the complete collinear kinematics using the 
definition of the parton momentum fraction as the ratio of light-cone components of the
parton and the nucleon \cite{Jaffe:1985je}. We show that there exist other
solutions for the parton momentum fractions $x_i$ which are neglected in the standard
parton model right from the start. These other solutions will turn
out to be unphysical and are derived at this point only to provide insight into
difficulties arising from a transverse-momentum dependent calculation as discussed in
Sec. \ref{subsubsec:naive}.

The partons inside the nucleons carry some fraction of their parent hadron's longitudinal momentum.
Labeling the parton momentum inside nucleon $i$ with $p_i$ we can define these fractions as ratios
of plus or minus components of the partons and the corresponding components of the nucleon 
momenta. In the Drell-Yan scaling limit ($S \rightarrow \infty$ and $M^2/S$ finite) 
$P_1^- = P_2^+ = \sqrt{S}$ become the large components while all other components vanish.
Note here that with a finite nucleon mass the large components would be modified
and the small components would be nonzero. This however poses no problem for the following
calculations, c.f. the discussion below Eqs. (\ref{nucleon1}) and (\ref{nucleon2}).
We define
\begin{align}
	x_1 &= \frac{p_1^-}{P_1^-} = \frac{p_1^-}{\sqrt{S}} \ ,     \label{x1} \\
        \tilde{x}_1 &= \frac{p_1^+}{\sqrt{S}}               \ ,     \label{x1_tilde} \\
	x_2 &= \frac{p_2^+}{P_2^+} = \frac{p_2^+}{\sqrt{S}} \ ,      \label{x2} \\
       	\tilde{x}_2 &= \frac{p_2^-}{\sqrt{S}} \ .		    \label{x2_tilde}	
\end{align}
Note that Eqs.\ (\ref{x1}) and (\ref{x2}) are standard definitions \cite{Jaffe:1985je}. The
tilde quantities in Eqs.\ (\ref{x1_tilde}) and (\ref{x2_tilde}) are introduced for later
convenience. The kinematical constraints for these fractions are the on shell conditions
\begin{eqnarray}
	p_1^2 &= p_1^+ p_1^- = 0 \Rightarrow x_1 \tilde{x_1} &= 0 \ , \label{onshell_cond1}   \\ 
	p_2^2 &= p_2^+ p_2^- = 0 \Rightarrow x_2 \tilde{x_2} &= 0 \ , \label{onshell_cond2}
\end{eqnarray}	
together with
\begin{align}
	M^2 = (p_1 + p_2)^2 = 2 p_1 p_2 &= p_1^+ p_2^- + p_1^- p_2^+ \nonumber \\
	    &= \left( \tilde{x_1} \tilde{x_2} + x_1 x_2 \right) S \label{invmass_cond}
\end{align}	    
and
\begin{align}
	x_F = \frac{\left( p_1 + p_2 \right)_z}{(q_z)_\textrm{max}}
	    &= \frac{1}{2(q_z)_\textrm{max}} \left( p_1^+ - p_1^- + p_2^+ - p_2^- \right)
		\nonumber \\
	    &= \frac{\sqrt{S}}{2 (q_z)_\textrm{max}} \left( \tilde{x_1} - x_1 + x_2 - \tilde{x_2}
			    \right)  \label{feynx_cond}
	    \ .
\end{align}
We will show now that the constraints in Eqs.\ (\ref{onshell_cond1})-(\ref{feynx_cond})
can be fulfilled by two different sets of momentum fractions $x_i, \tilde{x_i}$. 
Equation (\ref{onshell_cond1}) implies $\tilde{x_1} = 0$ or $x_1 = 0$.
If
\begin{align}      
      \tilde{x_1} &= 0 \\
      \xRightarrow{\textrm{Eq. (\ref{invmass_cond})} } \frac{M^2}{S} &= x_1 x_2 \label{spm_M2}\\
      \Rightarrow x_1 &\neq 0 \neq x_2 \\
      \xRightarrow{\textrm{Eq. (\ref{onshell_cond2})} } \tilde{x_2} &= 0 \\
      \xRightarrow{\textrm{Eq. (\ref{feynx_cond})} } x_F &= (x_2 - x_1)
                                 \frac{\sqrt{S}}{2 (q_z)_\textrm{max}} \label{spm_xF} .
\end{align}
This is just the standard parton-model solution, Eqs.\ (\ref{M2_pm},\ref{xF_pm}), as described
in Sec. \ref{subsec:spm}. However there exists another solution, namely for $x_1 = 0$:
\begin{align}      
      x_1 &= 0 \label{npm_x1} \\
      \xRightarrow{\textrm{Eq. (\ref{invmass_cond})}} \frac{M^2}{S} &= \tilde{x_1} \tilde{x_2} \\
      \Rightarrow \tilde{x_1} &\neq 0 \neq \tilde{x_2} \\
      \xRightarrow{\textrm{Eq. (\ref{onshell_cond2})}} x_2 &= 0 \\
      \xRightarrow{\textrm{Eq. (\ref{feynx_cond})}} x_F &= (\tilde{x_1} - \tilde{x_2})
				  \frac{\sqrt{S}}{2 (q_z)_\textrm{max}} \label{npm_xF}\ .
\end{align}
Kinematically this second solution represents the (strange) case where each parton moves into
the opposite direction of its respective parent nucleon. One can see this in the following 
example, where we choose $x_F = 0$. Then we have
\begin{align}
	\tilde{x_1} &= \tilde{x_2} = \frac{M}{\sqrt{S}}  \label{x1tilde_x2tilde} \\
	\Rightarrow p_1^z &= \frac{1}{2} \left( p_1^+ - p_1^- \right) 
			  = \frac{1}{2} \sqrt{S} \tilde{x_1} 
		  	  = \frac{M}{2} 
\end{align}
and analogously
\begin{equation}
	p_2^z = -\frac{M}{2}\ .
\end{equation}
Since nucleon 1 (2) moves into negative (positive) $z$-direction, cf.\ Eqs.\ (\ref{nucleon1}) and
(\ref{nucleon2}), the partons here move exactly {\it opposite}. 
The parton momentum fractions $x_i$ (not $\tilde x_i$!) entering the PDFs in Sec. \ref{subsec:spm} however are
those of partons that move into the {\it same} direction as their parent nucleon.
The second solution is thus
physically not meaningful and it is discarded right away in the standard parton model approach. 

The essential difference between the standard and the naive parton model is the following:
In the (collinear) standard parton model 
{\em all} components of $p_i$ are fixed at once by
$p_i = x_i P_i$. This automatically implies $\tilde x_i = 0$. Such a procedure is without
problems if one sticks to the collinear dynamics. In Sec. \ref{sec:full} below, however,
we include primordial transverse momenta of the partons, i.e.\ we have to deviate from
$p_i = x_i P_i$. The natural choice would be to define $x_i$ via one nucleon momentum component
(the large one). This is exactly what we have done here for the collinear case. 
In the naive parton model $x_i$ and $\tilde x_i$, i.e.\ $p_i^+$ and $p_i^-$, are introduced as
independent variables which are then constrained by the kinematical and onshell conditions
(\ref{onshell_cond1})-(\ref{feynx_cond}). However in the
Bjorken limit ($M,S \rightarrow \infty, M^2/S = \textrm{const}$) the parton momenta should
behave like \cite{Arnold:2008kf}

\begin{align}
	 p_1^- = O(M),\ \  & p_1^+ = O(1/M), \\
	 p_2^+  = O(M),\ \  & p_2^- = O(1/M) .
\end{align}	
For $\tilde x_i \neq 0$ this power counting is {\it not} fulfilled, cf. Eqs. (\ref{x1_tilde}) and
(\ref{x2_tilde}). Hence this solution corresponds 
to non-factorizing power suppressed corrections.
Therefore in the naive parton model one 
falls into a trap by picking up this additional unphysical solution. The same happens for the
more complicated case including primordial transverse momenta.

It is worth pointing out the connection between the two types of parton models (standard vs.
naive) and QCD. There, e.g., the DY cross section formula emerges from 
factorisation. It turns out that in the Bjorken limit a PDF depends on one variable only 
\cite{Jaffe:1985je}, which is encoded in $x_i$. In the final formula the energy-momentum
relation, e.g., for the DY process takes the form $\delta(q - x_1 P_1 - x_2 P_2)$ which suggests
the interpretation of $x_i P_i$ as the parton four-momentum. Thus the standard (collinear)
parton model emerges from QCD and not the naive one.

Including in addition primordial transverse momenta one has to model the distributions of
these momenta. However there is a constraint the chosen model has to
obey: in the Bjorken limit one should come back to the standard parton model and not to the 
naive one since only the former emerges from QCD.

In the following we will point out how to modify the naive parton model such that one ends up
with the standard parton model. This procedure will then be generalized to the case where
primordial transverse momenta of the partons are included.
In the naive parton model the hadronic cross section reads:
\begin{widetext}
\begin{align}
\frac{\diffd\sigma_{\textrm{naive}}}{\diffd M^2\ \diffd x_F}
         	&= \int_0^1 \textrm{d}x_1 \int_0^1 \textrm{d}x_2
		                  \sum_i q_i^2 f_i(x_1,M^2)\, f_{\bar i}(x_2,M^2)\,
				  \frac{4 \pi \alpha^2}{9 M^2} \, 
				  \delta\left(M^2 - (p_1+p_2)^2\right)\,
				  \delta\left(x_F - \frac{(p_1 + p_2)_z}{(q_z)_\textrm{max}}\right)
				  \nonumber \\
	        &= \int_0^1 \textrm{d}x_1 \int_0^1 \textrm{d}x_2
		                  \sum_i q_i^2 f_i(x_1,M^2)\, f_{\bar i}(x_2,M^2)\,
				  \frac{4 \pi \alpha^2}{9 M^2} \, 
				  \frac{2 (q_z)_\textrm{max} x_1 x_2
				        \left(\delta(x_1 - x_{1_+})\, \delta(x_2 - x_{2_+}) 
					 + \delta(x_1)\, \delta(x_2) \right)}
                                       {S^{3/2} \left| (x_1 x_2 -\tilde x_1 \tilde x_2)
			                       (x_1 + x_2 + \tilde x_1 + \tilde x_2)\right| }
\end{align}
\end{widetext}
The unphysical second solution for the momentum fractions is represented by 
\begin{equation} 
\delta(x_1)\, \delta(x_2)\, x_1 f_i(x_1,M^2)\, x_2 f_{\bar i}(x_2,M^2)
\end{equation}
in the last expression. Its contribution does not vanish since one obtains
for large enough $M^2$ \cite{Gluck:1998xa}
\begin{equation}
\underset{x\rightarrow 0}{\lim}\  \left(x\, f(x,M^2) \right) > 0\ . \label{lowx_pdf_limit}
\end{equation}

We now introduce a notation which we will keep throughout
this paper. Whenever we explicitly disregard unphysical solutions of the type of Eqs.\ 
(\ref{npm_x1})-(\ref{npm_xF}) under an integral we denote this integral by $\fint$. Thus
\begin{widetext}
\begin{align}
\frac{\diffd\sigma_{\textrm{naive}}}{\diffd M^2\ \diffd x_F}
         	&= \int_0^1 \textrm{d}x_1 \int_0^1 \textrm{d}x_2
		                  \sum_i q_i^2 f_i(x_1,M^2)\, f_{\bar i}(x_2,M^2)\, 
				  \frac{4 \pi \alpha^2}{9 M^2} \, 
				  \frac{2 (q_z)_\textrm{max} x_1 x_2
				        \left(\delta(x_1 - x_{1_+})\, \delta(x_2 - x_{2_+}) 
					 + \delta(x_1)\, \delta(x_2) \right)}
                                       {S^{3/2} \left| (x_1 x_2 -\tilde x_1 \tilde x_2)
			                       (x_1 + x_2 + \tilde x_1 + \tilde x_2)\right| }
\end{align}
wheras
\begin{align}			
	       	\frac{\diffd\sigma}{\diffd M^2\ \diffd x_F}
         	&= \fint_0^1 \textrm{d}x_1 \fint_0^1 \textrm{d}x_2
		                  \sum_i q_i^2 f_i(x_1,M^2)\, f_{\bar i}(x_2,M^2)\,
				  \frac{4 \pi \alpha^2}{9 M^2} \, 
				  \frac{2 (q_z)_\textrm{max} x_1 x_2
				        \left(\delta(x_1 - x_{1_+})\, \delta(x_2 - x_{2_+}) 
					 + \delta(x_1)\, \delta(x_2) \right)}
                                       {S^{3/2} \left| (x_1 x_2 -\tilde x_1 \tilde x_2)
			                       (x_1 + x_2 + \tilde x_1 + \tilde x_2)\right| }
				  \nonumber \\
		&= \int_0^1 \textrm{d}x_1 \int_0^1 \textrm{d}x_2
		                  \sum_i q_i^2 f_i(x_1,M^2)\, f_{\bar i}(x_2,M^2)\,
				  \frac{4 \pi \alpha^2}{9 M^2} \, 
				  \frac{2 (q_z)_\textrm{max} (x_1 x_2)
				        \delta(x_1 - x_{1_+})\, \delta(x_2 - x_{2_+})}
                                       {S^{3/2} (x_1 x_2) (x_1 + x_2) }
				   \nonumber \\    
		&= \sum_i q_i^2 f_i(x_{1_+},M^2)\, f_{\bar i}(x_{2_+},M^2)\,
				  \frac{4 \pi \alpha^2}{9 M^2} \, 
				  \frac{2 (q_z)_\textrm{max}}{S^{3/2} (x_{1_+} + x_{2_+}) }
				   \nonumber \\    
		&= \sum_i q_i^2 f_i(x_{1_+},M^2)\, f_{\bar i}(x_{2_+},M^2)\,
				  \frac{4 \pi \alpha^2}{9 M^2} \, 
				  \frac{(q_z)_\textrm{max}}
                                       {S E_\textrm{coll} }\ .
\end{align}
\end{widetext}
Note that in the last expression we have recovered the standard parton-model result
Eq.\ (\ref{spm_hadr_xsec}).

The main reason to present this naive approach in detail will become clear in the next section where we lift the 
simplification of a collinear movement of the partons with the nucleons.

\section{Full kinematics}
\label{sec:full}
The Bjorken limit and the corresponding infinite-momentum frame in which the standard parton model
is well defined and derived from leading-order pQCD is an
idealization of real experiments. There the nucleons will always move with some finite
momentum and thus the
partons inside the nucleons can have nonvanishing momentum components perpendicular to
the beam line. The factorisation into hard (subprocess) and soft (PDFs) physics 
is proven in the collinear case at least for leading twist (expansion in $1/M$) in
\cite{Collins:1989gx} and in the transverse case at least for partons with low transverse 
momentum in \cite{Ji:2004xq}.

\subsection{Transverse-momentum distributions}
\label{subsec:tmd}

For the calculation of the hadronic cross sections we will need transverse-momentum dependent
parton distribution functions. We denote these by ${\tilde f}_i$. They  
are functions of the light-cone momentum fraction $x_i$, the transverse 
momentum ${\vec p}_{i_\perp}$ and the hard scale of the subprocess $q^2$. The general form of these
functions however is unknown. Known rather well are the longitudinal PDFs.
Since data of DY pair production are compatible with a Gaussian form of the $p_T$-spectrum up 
to a certain $p_T$ \cite{Webb:2003bj,Webb:2003ps}, we assume factorisation of the longitudinal 
and the transverse part of ${\tilde f}_i$ and make the following common ansatz 
\cite{Wang:1998ww,Raufeisen:2002zp,D'Alesio:2004up}
\begin{equation}
   {\tilde f}_i(x_i,{\vec p}_{i_\perp},q^2) = f_i(x_i,q^2) \cdot f_{i_\perp}({\vec p}_{i_\perp})\ .
   \label{ftilde}
\end{equation}	
Here $f_i$ are the usual longitudinal PDFs and for $f_{i_\perp}$ we choose a Gaussian form,
\begin{equation}
	f_{i_\perp}({\vec p}_{i_\perp}) = \frac{1}{4\pi D^2} 
				\exp\left( -\frac{({\vec p}_{i_\perp})^2} {4 D^2}\right) \ .
	\label{fperp}
\end{equation}	
The width parameter $D$ is connected to the average squared transverse momentum via
\begin{equation}
	\left< ({\vec p}_{i_\perp})^2 \right> = 
	\int \diffd {\vec p}_{i_\perp} ({\vec p}_{i_\perp})^2 f_{i_\perp}({\vec p}_{i_\perp})
	= 4 D^2 \ 
\end{equation}
and it has to be fitted to the available data.

\subsection{Cross section}
\label{subsec:fullsol}

Now we calculate the hadronic cross section $\diffd\sigma$ taking into account the full
kinematics. Since it is necessary to remove the unphysical solutions for the light-cone momentum
fractions $x_1$ and $x_2$ which correspond to the ones found in Sec. \ref{subsec:npm} 
for the collinear case, the calculation has to be conducted such that it is possible
to disentangle the physical and the unphysical solution. 
First we will discuss in Sec. \ref{subsubsec:naive} a straightforward calculation 
which however does {\it not} obey this requirement (the details of this calculation can be
found in the Appendix). If one does not remove the unphysical
solutions one produces unphysical results. This reveals a pitfall which the unawareness of this
problem can create \cite{Linnyk:2004mt,Linnyk:2005iw,Linnyk:2006mv}. In Sec.
\ref{subsubsec:correct} we will show how to properly remove the unphysical solutions as we did 
for the collinear case at the end of Sec. \ref{subsec:npm}.

In the transverse-momentum dependent approach the leading-order Drell-Yan total differential
cross section reads \cite{D'Alesio:2004up}
\begin{widetext}
\begin{equation}
	\textrm{d}\sigma = \int_0^1 \textrm{d}x_1 \int_0^1 \textrm{d}x_2
       		           \int \textrm{d}{\vec p}_{1_\perp} \int \textrm{d}{\vec p}_{2_\perp}
		           \sum_i q_i^2 {\tilde f}_i(x_1,{\vec p}_{1_\perp},q^2)
				{\tilde f}_{\bar i}(x_2,{\vec p}_{2_\perp},q^2) \cdot
			   \textrm{d}\hat\sigma(x_1,{\vec p}_{1_\perp},x_2,{\vec p}_{2_\perp},q^2)
			   \ .        \label{hadr_full_xsec}
\end{equation}
\end{widetext}
In this approach the transverse momentum ($p_T=|{\vec q}_\perp|$) of the DY pair is accessible,
since the annihilating quark and antiquark can have finite initial transverse momenta.
Note that in the calculations of \cite{D'Alesio:2004up} the partonic DY cross section 
$\diffd \hat \sigma$ was taken in the collinear limit.

\subsubsection{Naive calculation}
\label{subsubsec:naive}

In the naive approach the partonic triple-differential cross section reads:
\begin{widetext}
\begin{align}
	\frac{\textrm{d}\hat \sigma_\textrm{naive}}{\diffd M^2\ \diffd x_F \diffd p_T^2} 
				&= 
				\int \diffd^4q
                                \frac{4 \pi \alpha^2}{9 q^2} \, \delta\left(M^2 - q^2\right) \,
				\delta^{(4)}(p_1 + p_2 - q) \, 
				\delta\left(x_F - \frac{q_z}{(q_z)_\textrm{max}}\right)
				\delta\left(p_T^2 - ({\vec q}_\perp)^2 \right)
								\nonumber   \\
                             	&= 
                                \frac{4 \pi \alpha^2}{9 M^2} \, 
				\delta\left(M^2 - (p_1+p_2)^2\right)\,
				\delta\left(x_F - \frac{(p_1 + p_2)_z}{(q_z)_\textrm{max}}\right)
				\delta\left(p_T^2 - ({\vec p}_{1_\perp} + {\vec p}_{2_\perp})^2
											\right)
				\ .
				\label{triplepart_naive_xsec}
\end{align}
\end{widetext}
Inserting Eq.\ (\ref{triplepart_naive_xsec}) in Eq.\ (\ref{hadr_full_xsec}) yields a multiple 
integral for the triple-differential cross section:
\begin{widetext}
\begin{align}
	\frac{\textrm{d} \sigma_\textrm{naive}}{\diffd M^2\ \diffd x_F \diffd p_T^2} 
	                   = \int_0^1 \textrm{d}x_1 \int_0^1 \textrm{d}x_2
	                   \int \textrm{d}{\vec p}_{1_\perp} \int \textrm{d}{\vec p}_{2_\perp}
		           &F(x_1,{\vec p}_{1_\perp},x_2,{\vec p}_{2_\perp},M^2)\, \nonumber   \\ 
			   \times\ &\delta\left(M^2 - (p_1+p_2)^2\right)\,
			   \delta\left(x_F - \frac{(p_1 + p_2)_z}{(q_z)_\textrm{max}}\right)
			   \delta\left(p_T^2 - ({\vec p}_{1_\perp} + {\vec p}_{2_\perp})^2 \right)
			   \ .
			\label{naive_full_integral}
\end{align}
\end{widetext}
All pieces which do not contain $\delta$-functions are collected in $F(...)$.
The straightforward, but naive calculation of (\ref{naive_full_integral}) was performed in
\cite{Linnyk:2004mt,Linnyk:2005iw,Linnyk:2006mv}. The details of this calculation can be found in the
Appendix. Here we just want to point out the problems arising from this approach:
the naive calculation with the full kinematics incorporates unphysical solutions for the
momentum fractions $x_i$ which correspond to the unphysical solutions of the collinear case
in Eqs.\ (\ref{npm_x1})-(\ref{npm_xF}). However in the collinear kinematics it is quite clear that
these solutions for $x_i$ {\em cannot} be the physically interesting ones, since they are just
$x_i = 0$
and the PDFs are divergent for small $x$ as one can conclude from (\ref{lowx_pdf_limit}).
In the
case of full kinematics the situation is similar, however due to the introduction of transverse
quark momentum distributions the momentum fractions $x_i$ are smeared out around their collinear
values. Nonetheless the unphysical solutions are still very close to zero and one picks up
very large contributions of the diverging PDFs at such low $x$. This leads to a large enhancement
of the cross section in the full kinematical approach and the data are now {\em overestimated}.
The effect can be seen in \cite{Linnyk:2004mt,Linnyk:2005iw,Linnyk:2006mv} and also in Sec. 
\ref{sec:res} where we compare this naive approach and the correct calculation of the next section.
In addition, in Sec. \ref{sec:res} it will be shown that the $M$-dependence of the cross 
section is {\em not} reproduced in the naive approach.

\subsubsection{Correct calculation}
\label{subsubsec:correct}

The analogue to Eq.\ (\ref{naive_full_integral}) in the correct approach is
(for the notation $\fint$ see Sec. \ref{subsec:npm}) 
\begin{widetext}	
\begin{align}
	\frac{\textrm{d} \sigma}{\diffd M^2\ \diffd x_F \diffd p_T^2} 
	                   = \fint_0^1 \textrm{d}x_1 \fint_0^1 \textrm{d}x_2
	                   \int \textrm{d}{\vec p}_{1_\perp} 
			   \int \textrm{d}{\vec p}_{2_\perp}
		           &F(x_1,{\vec p}_{1_\perp},x_2,{\vec p}_{2_\perp},M^2)\, \nonumber   \\ 
			   \times\, &\delta\left(M^2 - (p_1+p_2)^2\right)\,
			   \delta\left(x_F - \frac{(p_1 + p_2)_z}{(q_z)_\textrm{max}}\right)\,
			   \delta\left(p_T^2 - ({\vec p}_{1_\perp} + {\vec p}_{2_\perp})^2 \right)
			   \ .
			\label{correct_full_integral}
\end{align}
\end{widetext}
The $\delta$-functions in Eq.\ (\ref{correct_full_integral}) must be worked out in a way 
that allows to discern physical and unphysical solutions for the momentum fractions $x_i$ in order
to perform the $\fint$-integrations. For this aim it is useful to rewrite the parton 
momenta in terms of different variables:
\begin{align}
	q &=  p_1 + p_2 \ , \\
	k &= \frac{1}{2} \left( p_2 - p_1 \right) \ .
\end{align}
Inverting the last two equations, we can use the on-shell conditions for the partons to get
\begin{equation}
	0 = p_1^2 
	  = \left(\frac{1}{2} q - k \right)^2      
	  = \frac{1}{4} q^2 - k \cdot q + k^2 \label{onshell_cond1_full}
\end{equation}	  
and
\begin{equation}
	0 = p_2^2 
	  = \left(\frac{1}{2} q + k \right)^2       
	  = \frac{1}{4} q^2 + k \cdot q + k^2  \label{onshell_cond2_full} \ .
\end{equation}	  
Adding and subtracting Eqs.\ (\ref{onshell_cond1_full}) and (\ref{onshell_cond2_full})
yields
\begin{align}
	k^2      &= -\frac{1}{4} M^2 \ , \label{ksquare} \\
	k\cdot q  &= 0		     \label{kq}	\ .
\end{align}	
Solving Eq.\ (\ref{ksquare}) for $k^+$ yields
\begin{equation}
	k^+ = \frac{\vec k_{\perp}^2 - \frac{1}{4}M^2}{k^-} \ .
\end{equation}
Inserting this result into Eq.\ (\ref{kq}) gives an equation quadratic in $k^-$:
\begin{align}
	0 &= k^+ q^- + k^- q^+ - 2 \vec k_{\perp} \cdot \vec q_{\perp}  \nonumber    \\
	  &= \frac{\vec k_{\perp}^2 - \frac{1}{4}M^2}{k^-} q^- + k^- q^+ 
	                                             - 2 \vec k_{\perp} \cdot \vec q_{\perp} 
						     \label{kq_comp}  \\
        \Rightarrow 0 &= \left (k^-\right)^2 q^+ - 2 \vec k_{\perp} \cdot \vec q_{\perp} k^- 
                                         + \left( \vec k_{\perp}^2 - \frac{1}{4}M^2 \right)q^- \ .
\end{align}	
The solutions are
\begin{equation}
	(k^-)_\pm = \frac{\vec k_{\perp} \cdot \vec q_{\perp}}{q^+} \pm 
	      \sqrt{ \left( \frac{\vec k_{\perp} \cdot \vec q_{\perp}}{q^+} \right)^2 
	       + \frac{q^-}{q^+} \left(\frac{1}{4} M^2 - \vec k_{\perp}^2 \right) } \ .
	       \label{kminus_sol}
\end{equation}	
Inserting (\ref{kminus_sol}) into (\ref{kq_comp}) gives the solutions for $k^+$:
\begin{equation}
	(k^+)_\mp = \frac{q^+}{q^-} \left( \frac{\vec k_{\perp} \cdot \vec q_{\perp}}{q^+} \mp 
		    \sqrt{ \left( \frac{\vec k_{\perp} \cdot \vec q_{\perp}}{q^+} \right)^2 
		    + \frac{q^-}{q^+} \left(\frac{1}{4} M^2 - \vec k_{\perp}^2 \right)}\  \right)
		\ .
  		\label{kplus_sol}
\end{equation}
Rewriting now Eqs.\ (\ref{x1}) and (\ref{x2}) in terms of $q$ and $k$ we obtain the solutions
for the parton momentum fractions:
\begin{align}
	(x_1)_\pm = \frac{p_1^-}{\sqrt{S}} 
	    = \frac{1}{\sqrt{S}} \left( \frac{1}{2} q^- - (k^-)_\pm \right) \label{x1_full}
\end{align}
and
\begin{align}
	(x_2)_\mp = \frac{p_2^+}{\sqrt{S}} 
	    = \frac{1}{\sqrt{S}} \left( \frac{1}{2} q^+ + (k^+)_\mp \right) \label{x2_full} \ .
\end{align}
Since there are two solutions for $k^-$ and $k^+$, respectively, we also get two solutions
for $x_1$, $x_2$. To determine which set of $x_1,x_2$ and thus $k^+,k^-$ has to be chosen we take
the limit of zero parton transverse momentum. In this way one can make the connection to the 
collinear case (then $q^2 \rightarrow q^+ q^- = M^2$):
\begin{eqnarray}
	(k^-)_\pm &\rightarrow \pm \sqrt{\frac{q^-}{q^+} \frac{1}{4} M^2} &= \pm \frac{q^-}{2} 
									\label{kminus_coll} \\
	(k^+)_\mp &\rightarrow \mp \sqrt{\frac{q^+}{q^-} \frac{1}{4} M^2} &= \mp \frac{q^+}{2} 
									\label{kplus_coll}
\end{eqnarray}	
Inserting expressions (\ref{kminus_coll}) and (\ref{kplus_coll}) into (\ref{x1_full}) and (\ref{x2_full}) 
yields two solutions for the momentum fractions, just as in the collinear case in Sec. 
\ref{subsec:npm}:
\begin{equation}
	(x_1)_\pm \rightarrow \frac{1}{\sqrt{S}} \begin{cases} 
					0 \\
					q^-
				  \end{cases} 
\end{equation}				 
and
\begin{equation}
	(x_2)_\mp \rightarrow \frac{1}{\sqrt{S}} \begin{cases}
					0 \\
					q^+
				  \end{cases} \ .
\end{equation}			  	  
The lower solutions correspond to the standard parton model Eqs.\ (\ref{spm_M2}), (\ref{spm_xF}),
since $x_1 x_2 = \frac{M^2}{S}$ and $x_2 - x_1 = \frac{2 q_z}{\sqrt{S}} = x_F 
\frac{2 (q_z)_{\textrm{max}}}{\sqrt{S}}$. The upper
solution then corresponds to the unphysical case $x_1 = x_2 = 0$ and $\tilde x_1 \neq 0 \neq
\tilde x_2$, see Eqs.\ (\ref{npm_x1})-(\ref{npm_xF}).

This is the crucial point: to receive physically meaningful results from Eq. 
(\ref{correct_full_integral})
one has to discard these upper solutions just as one does in the collinear case in Sec.
\ref{subsec:npm}. This requires that the integrals in Eq. (\ref{correct_full_integral}) are 
evaluated in the correct order, otherwise one cannot disentangle the two different 
solutions for $x_1$
and $x_2$. We will now present a calculation which respects this requirement. In Sec. 
\ref{sec:res} we will show that the quantitative difference between this calculation and the
calculation from Sec. \ref{subsubsec:naive} is huge.

We begin by introducing several integrals over $\delta$-functions in Eq.\  
(\ref{correct_full_integral}).
In this way we will transform the integration variables to the above chosen $q$ and $\vec k_\perp$:
\begin{widetext}
\begin{align}
	\frac{\textrm{d} \sigma}{\diffd M^2\ \diffd x_F \diffd p_T^2} 
	                   = &\fint_0^1 \textrm{d}x_1 \fint_0^1 \textrm{d}x_2
	                   \int \textrm{d}{\vec p}_{1_\perp} 
			   \int \textrm{d}{\vec p}_{2_\perp}
			   \int \textrm{d}{\vec q}_{\perp}
			   \int \textrm{d}{\vec k}_{\perp}
			   \int\diffd q^+ \int \diffd q^- 
			   F(x_1,{\vec p}_{1_\perp},x_2,{\vec p}_{2_\perp},M^2) \nonumber \\ 
		           &\times \delta(q^+ - (p_1^+ + p_2^+))\, \delta(q^- - (p_1^- + p_2^-))\,
			   \delta^{(2)}({\vec q}_{\perp} - \left({\vec p}_{1_\perp} 
						         + {\vec p}_{2_\perp} \right))\,
			   \delta^{(2)}({\vec k}_{\perp} - \frac{1}{2}\left({\vec p}_{1_\perp} 
						                      - {\vec p}_{2_\perp} \right))
			   \nonumber \\
			   &\times \delta\left(M^2 - (p_1+p_2)^2\right)\,
			   \delta\left(x_F - \frac{(p_1 + p_2)_z}{(q_z)_\textrm{max}}\right)
			   \delta\left(p_T^2 - ({\vec p}_{1_\perp} + {\vec p}_{2_\perp})^2 \right)
			\label{full_integral_correct} \ .
\end{align}
First we perform
\begin{equation}
	\int \textrm{d}{\vec p}_{1_\perp} \int \textrm{d}{\vec p}_{2_\perp}
	\delta^{(2)}\left({\vec q}_{\perp} - \left({\vec p}_{1_\perp} + {\vec p}_{2_\perp} \right)
		    \right)\,
	\delta^{(2)}\left({\vec k}_{\perp} - \frac{1}{2}\left({\vec p}_{2_\perp} - {\vec p}_{1_\perp} 
				\right)\right)
	= 1 \ .
\end{equation}	
\end{widetext}
Now we calculate the integral
\begin{align}
	\fint_0^1 \textrm{d}x_1 \fint_0^1 \textrm{d}x_2\  
	 \delta\left(q^+ - (p_1^+ + p_2^+)\right)\, \delta\left(q^- - (p_1^- + p_2^-)\right)\ . 
\end{align}       
According to Eqs.\ (\ref{kminus_sol})-(\ref{x2_full}) the $\delta$-functions in the last expression
have two possible solutions for each $p_1^-$ and $p_2^+$. However as explained above
we now have to explicitly remove the unphysical solutions $(x_1)_+$ and $(x_2)_-$ , which are the ones
corresponding to the upper sign in Eqs.\ (\ref{kminus_sol}) and (\ref{kplus_sol}):
\begin{widetext}
\begin{align}
	&\fint_0^1 \textrm{d}x_1 \fint_0^1 \textrm{d}x_2\  
	 \delta(q^+ - (p_1^+ + p_2^+))\, \delta(q^- - (p_1^- + p_2^-)) \nonumber \\
       =&\fint_0^1 \textrm{d}x_1 \fint_0^1 \textrm{d}x_2\  
	 \delta\left(q^+ 
	       - \frac{\left(\frac{1}{2}{\vec q}_{\perp} - {\vec k}_{\perp}\right)^2}{x_1\sqrt{S}}
	       - x_2 \sqrt{S} \right)\,
	 \delta\left(q^- - x_1 \sqrt{S} 
               - \frac{\left(\frac{1}{2}{\vec q}_{\perp} + {\vec k}_{\perp}\right)^2}{x_2\sqrt{S}}
	       \right)
		 \nonumber \\
       =&\int_0^1 \textrm{d}x_1 \int_0^1 \textrm{d}x_2\  
	 \delta\left(q^+ - \frac{\left(\frac{1}{2}{\vec q}_{\perp} - {\vec k}_{\perp}\right)^2}
			        {(x_1)_-\sqrt{S}} 
	            - (x_2)_+ \sqrt{S}\right)\, \delta\left(q^- - (x_1)_- \sqrt{S} 
		    - \frac{\left(\frac{1}{2}{\vec q}_{\perp} + {\vec k}_{\perp}\right)^2}
		           {(x_2)_+\sqrt{S}}\right) 
	 \nonumber \\
       =&\left| S - \frac{(\frac{1}{2}{\vec q}_{\perp} - {\vec k}_{\perp})^2
			           (\frac{1}{2}{\vec q}_{\perp} + {\vec k}_{\perp})^2}
				  {(x_1)_-^2 (x_2)_+^2 S}\right|^{-1} .	
\end{align}       
Using $\diffd q^+ \diffd q^- = 2 \diffd q_0 \diffd q_z$ we can evaluate some of the remaining 
integrals of Eq.\ (\ref{full_integral_correct}) with the help of the $\delta$-functions:
\begin{align}
	&\int \diffd q^+ \diffd q^- \diffd {\vec q}_{\perp}
	    \delta\left(M^2 - q^2\right)\,
	    \delta\left(x_F - \frac{q_z}{(q_z)_\textrm{max}}\right)\,
	    \delta\left(p_T^2 - ({\vec q}_{\perp})^2 \right) \nonumber \\
       =2&\int \diffd q_0 \diffd {\vec q}_{\perp} \diffd q_z\ 
       	    \delta\left(M^2 + ({\vec q}_{\perp})^2 + q_z^2  - q_0^2\right)\,
	    \delta\left(x_F - \frac{q_z}{(q_z)_\textrm{max}}\right)\,
	    \delta\left(p_T^2 - ({\vec q}_{\perp})^2 \right) \nonumber \\
       =\ \ &\frac{\pi\, (q_z)_\textrm{max}}{E}
\end{align}	
with $E = \sqrt{M^2 + p_T^2 + x_F^2(q_z)_\textrm{max}^2}$\ .
Collecting the pieces, what remains of Eq.\ (\ref{full_integral_correct}) is
\begin{equation}
	\frac{\textrm{d} \sigma}{\diffd M^2\ \diffd x_F \diffd p_T^2} 
	                   =  \int^{|{\vec k}_{\perp}|_\textrm{max}} \textrm{d}{\vec k}_{\perp}
			       \frac{\pi\ (q_z)_\textrm{max}}{E}
				\left| S - \frac{(\hat {\vec p}_{1_\perp})^2
			           (\hat {\vec p}_{2_\perp})^2 }
				  {(x_1)_-^2 (x_2)_+^2 S}\right|^{-1}
			F((x_1)_-,\hat {\vec p}_{1_\perp},(x_2)_+,\hat {\vec p}_{2_\perp},M^2) 
		\ .
\end{equation}
$(x_1)_-,\hat {\vec p}_{1_\perp},(x_2)_+$ and $\hat {\vec p}_{2_\perp}$ are now fixed:
\begin{align}
	(x_1)_- &= \frac{1}{\sqrt{S}}\left(\frac{q^-}{2} 
			 - \frac{\vec k_{\perp} \cdot \vec q_{\perp}}{q^+} + 
	      		\sqrt{ \left( \frac{\vec k_{\perp} \cdot \vec q_{\perp}}{q^+} \right)^2 
	       	    + \frac{q^-}{q^+} \left(\frac{1}{4} M^2 - \vec k_{\perp}^2 \right) }\right)
											\ , \\
	(x_2)_+ &= \frac{1}{\sqrt{S}} \left(\frac{q^+}{2}
		         + \frac{\vec k_{\perp} \cdot \vec q_{\perp}}{q^-} 
			 + \sqrt{ \left( \frac{\vec k_{\perp} \cdot \vec q_{\perp}}{q^-} \right)^2 
		    + \frac{q^+}{q^-} \left(\frac{1}{4} M^2 - \vec k_{\perp}^2 \right) }\right)
											\ , \\
	\hat {\vec p}_{1_\perp}	&= \frac{1}{2}{\vec q}_{\perp} - {\vec k}_{\perp} \ , \\
	\hat {\vec p}_{2_\perp}	&= \frac{1}{2}{\vec q}_{\perp} + {\vec k}_{\perp}
\end{align}
with
\begin{align}
	q^+ &= E + x_F (q_z)_\textrm{max} \ , \\
	q^- &= E - x_F (q_z)_\textrm{max} \ , \\
	\left|{\vec q}_{\perp}\right| &= p_T \ ,\\
	\vec k_{\perp} \cdot \vec q_{\perp} &= |\vec k_{\perp}| 
	       p_T \cos(\phi_\perp) \ .
\end{align}
\end{widetext}
$|{\vec k}_{\perp}|_\textrm{max}$ is fixed by the condition that $(x_1)_-$ and $(x_2)_+$ 
must be real numbers:
\begin{equation}
	({\vec k}_{\perp})^2 < \frac{(M^2 + p_T^2) \frac{M^2}{4}}{M^2 + p_T^2(1-\cos^2(\phi_\perp))}
	= ({\vec k}_{\perp})^2_\textrm{max}\ 
\end{equation}
We have convinced ourselves that this condition also guarantees that $0 < (x_1)_-,(x_2)_+ < 1$.
Finally we arrive at the following expression:
\begin{widetext}
\begin{equation}
	\frac{\textrm{d} \sigma}{\diffd M^2\ \diffd x_F \diffd p_T^2} 
	                   = \int_0^{2\pi} \diffd \phi_\perp 
			   \int_0^{({\vec k}_{\perp})^2_\textrm{max}} \frac{1}{2}
			        \diffd ({\vec k}_{\perp})^2
			       \frac{\pi\ (q_z)_\textrm{max}}{E}
			   \left| S - \frac{(\hat {\vec p}_{1_\perp})^2
			           (\hat {\vec p}_{2_\perp})^2 }
				  {(x_1)_-^2 (x_2)_+^2 S}\right|^{-1}
			F((x_1)_-,\hat {\vec p}_{1_\perp},(x_2)_+,\hat {\vec p}_{2_\perp},M^2)
			  \label{master_correct}
\end{equation}
with
\begin{equation}
	F((x_1)_-,\hat {\vec p}_{1_\perp},(x_2)_+,\hat {\vec p}_{2_\perp},M^2)
	= \sum_i q_i^2\, {\tilde f}_i\left((x_1)_-,\hat {\vec p}_{1_\perp},M^2\right)\,
		       {\tilde f}_{\bar i}\left((x_2)_+,\hat {\vec p}_{2_\perp},M^2\right)\,
		       \frac{4 \pi \alpha^2}{9 M^2} 
\end{equation}	
and with ${\tilde f}_i$ defined in Eq.\ (\ref{ftilde}).
\end{widetext}

\section{Results}
\label{sec:res}

In this section we present our quantitative results and compare the naive approach
of Sec. \ref{subsubsec:naive} and the correct approach of Sec. \ref{subsubsec:correct}.
The data are from the NuSea Collaboration (E866) \cite{Webb:2003bj,Webb:2003ps} and
from FNAL-E439 \cite{Smith:1981gv}. For the collinear PDFs we used the GRV98 LO parametrization
\cite{Gluck:1998xa} available through CERN's PDFLIB version 8.04 \cite{Plothow:2000aa}.

\subsection{E866 -- $p_T$-spectra}
Experiment E866 measured continuum dimuon production in pp collisions at $S \approx 1500$ GeV$^2$.
The triple-differential cross section as given by the E866 collaboration is
\begin{equation}
	E\frac{\diffd \sigma}{d^3p} \equiv \frac{2E}{\pi \sqrt{S}} \frac{\diffd \sigma}
							     	   {\diffd x_F \diffd p_T^2}
        \label{E866_triple_xsec}								   
\end{equation}	
where an average over the azimuthal angle has been taken. The data are given in several
bins of $M$, $x_F$ and $p_T$ and for every datapoint the average values $\left< M \right>$,
$\left< x_F \right>$ and $\left< p_T \right>$ are given. Since our schemes provide 
Eqs.\ (\ref{master_naive}) and (\ref{master_correct}) we calculate the quantity of 
Eq.\ (\ref{E866_triple_xsec}) for every datapoint using these averaged values and 
then perform a simple average in every $M^2$-bin:
\begin{align}
	\frac{2E}{\pi \sqrt{S}} \frac{\diffd \sigma}{\diffd x_F \diffd p_T^2}
       &\rightarrow\frac{2E}{\pi \sqrt{S}} \int_{M^2\textrm{-bin}}
         \frac{\diffd \sigma}{\diffd M^2 \diffd x_F \diffd p_T^2} \diffd M^2 \nonumber \\
       &\approx 	\frac{2E}{\pi \sqrt{S}} \Delta M^2
                \frac{\diffd \sigma}{\diffd M^2 \diffd x_F \diffd p_T^2}\left(\left< M \right>,
									      \left< x_F \right>,
									      \left< p_T \right>
									\right)
									  \ ,
\end{align}
where
\begin{equation}
      E = \sqrt{\vrule height 10pt width 0pt 
	         \left< M \right>^2 + \left< p_T \right>^2 + 
	         \left< x_F \right>^2 \left<(q_z)_\textrm{max}\right>^2 }
\end{equation}
and $\Delta M^2 = M_\textrm{max}^2 - M_\textrm{min}^2$ with $M_\textrm{max}$ ($M_\textrm{min}$)
the upper (lower) limit of the bin. 	

\begin{figure}[H]
	\centering
	\includegraphics[angle=-90,keepaspectratio,width=0.45\textwidth]
                        {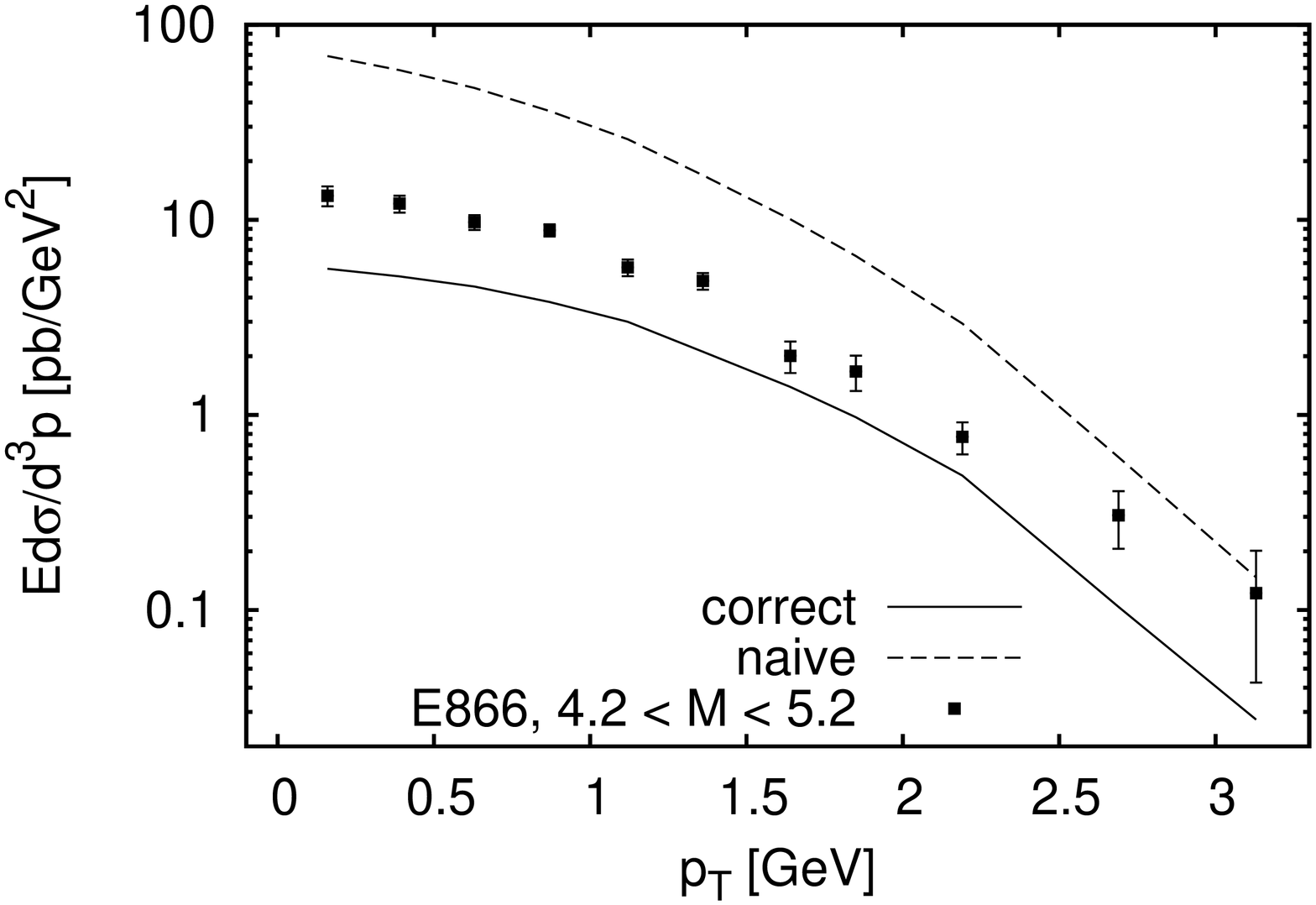}
	\includegraphics[angle=-90,keepaspectratio,width=0.45\textwidth]
                        {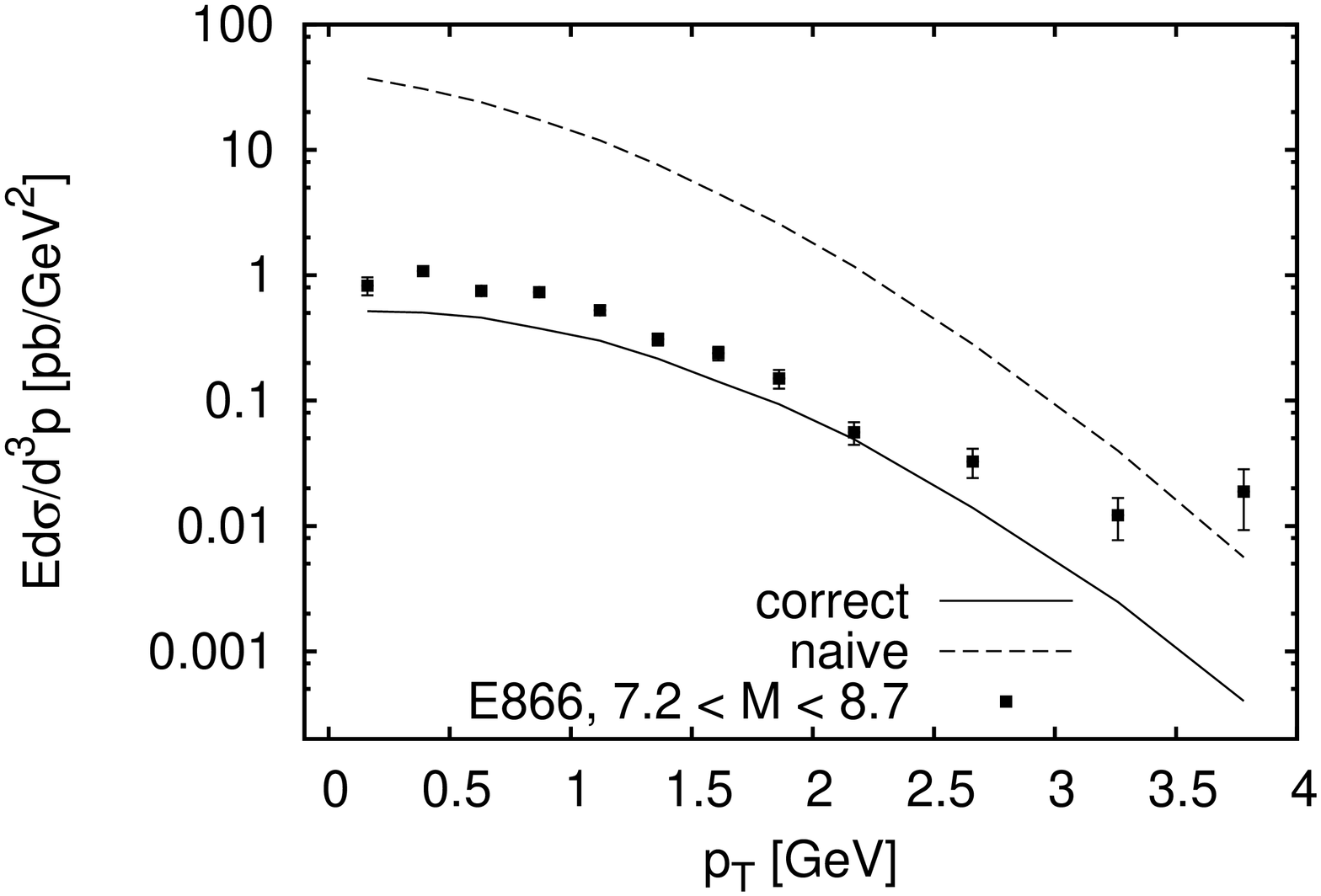}

	\caption{$p_T$-spectrum obtained from the naive and the correct approach with $D=0.5$ GeV. 
		 Data are from E866 binned with \\ $4.2$ GeV $< M < 5.2$ GeV and
		 $7.2$ GeV $< M < 8.7$ GeV,\\ $-0.05 < x_F < 0.15$.
		 Only statistical errors are shown.}
	\label{fig:E866_triple_4.2}	 
\end{figure}

We plot the results for the two different approaches in different $M$-bins in Fig.
\ref{fig:E866_triple_4.2}.
Everywhere a value of $D=0.5$ GeV for the transverse momentum dispersion was chosen.
The solid lines represent the correct approach. The shape of the spectra is described
rather well which is due to the choice of the parameter $D$. However the absolute
size is still underestimated and a factor $K\approx 1.75 - 2.0$ would be necessary to reproduce
the height of the data. 
The naive approach is plotted with dashes. As already mentioned in Sec. \ref{subsubsec:naive}
the calculated cross section overestimates the data significantly. This can also be seen
in \cite{Linnyk:2004mt,Linnyk:2005iw,Linnyk:2006mv}. We note that the discrepancy between 
both approaches is about 1 order of magnitude and it becomes worse in the higher
mass bin. This already indicates a wrong $M$ dependence of the naive approach.

\label{subsec:res_triple}

\subsection{E866 - $M$-spectrum}
\label{subsec:res_double}

The double-differential cross section is given by the E866 collaboration as
\begin{equation}
	M^3\frac{\diffd \sigma}{\diffd M \diffd x_F}\ .         \label{E866_double_xsec}								   
\end{equation}	
Again the data are given in several bins of $M$ and $x_F$ and for every datapoint the 
average values $\left< M \right>$ and $\left< x_F \right>$ are provided. 
Once more we start with Eqs.\ (\ref{master_naive}) and (\ref{master_correct}) and calculate the quantity
of Eq.\ (\ref{E866_double_xsec}) by integrating over $p_T^2$ for every datapoint using these 
averaged values:
\begin{widetext}
\begin{align}
	 M^3\frac{\diffd \sigma}{\diffd M \diffd x_F}
 	&\rightarrow \left<M\right>^3 \int_0^{(p_T)^2_\textrm{max}} \diffd p_T^2
	      \frac{\diffd \sigma}{\diffd M \diffd x_F \diffd p_T^2}
	=\left<M\right>^3 \int_0^{(p_T)^2_\textrm{max}} \diffd p_T^2\ 
	 2 \left<M\right> \frac{\diffd \sigma}{\diffd M^2 \diffd x_F \diffd p_T^2} 
           \left(\left< M \right>, \left< x_F \right> \right) \ .
\end{align}
\end{widetext}
The maximal possible $p_T$ is determined by the kinematics.
\begin{align}
	P_1 + P_2 &= q + X \\
	\Rightarrow (P_1 + P_2 - q)^2 &= X^2 = M_R^2 \\
	\Rightarrow S + M^2 - M_R^2 &= 2\ (P_1 + P_2)\ q \nonumber \\
	                            &= 2 \sqrt{S} E \nonumber \\
				    &= 2 \sqrt{S} \sqrt{M^2 + p_T^2 + q_z^2}\\
	\Rightarrow M^2 + (p_T)^2_\textrm{max} + q_z^2 &= E^2 = \frac{(S+M^2-M_R^2)^2}{4S} \\	    
	\Rightarrow (p_T)^2_\textrm{max} &= \frac{(S+M^2-M_R^2)^2}{4S} - M^2 - q_z^2 \ .
\end{align}	
$M_R^2$ is the minimal invariant mass of the undetected remnants. We choose a value
of $M_R=1.1$ GeV. Note that at c.m.\ energies of $\sqrt{S} \approx 27.4$ GeV (E439)
and $\sqrt{S} \approx 38.8$ GeV (E866) we are not really sensitive to this value if it 
stays at or below a few GeV.

\begin{figure}[H]
	\centering
	\includegraphics[angle=-90,keepaspectratio,width=0.45\textwidth]
                        {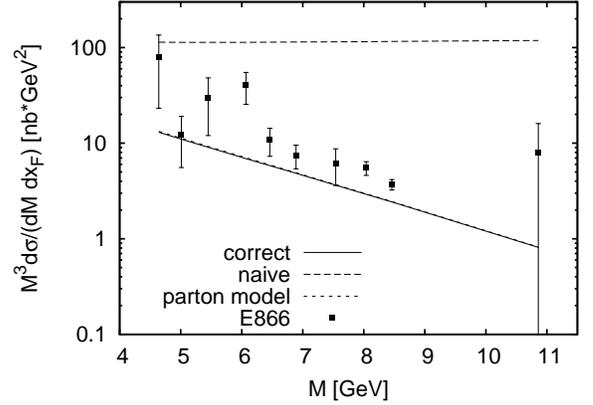}
	\caption{$M$-spectrum obtained from naive and correct approach with $D=0.5$ and
		 from the standard parton model. Note that the results from the latter
		 two approaches are basically on top of each other.	 
		 Data are from E866 binned with $-0.05 < x_F < 0.05$.
		 Only statistical errors shown.}
	\label{fig:E866_double}	 
\end{figure}

The results are plotted in Fig. \ref{fig:E866_double}. Again we use $D=0.5$ GeV. The solid line 
represents
the correct approach, the long dashed line the naive one. For comparison the result of the
standard (collinear) parton model is plotted with the short dashed line.
Here the discrepancy between the naive and the correct approach is
fully visible, since neither the slope nor the size of the $M$-spectrum is reproduced in the
naive approach. Instead it gives almost a constant distribution. (Note here that this dataset is not
shown or compared to calculations in \cite{Linnyk:2004mt,Linnyk:2005iw,Linnyk:2006mv}.)
The correct approach however
describes the slope well and again a factor $K\approx 1.75-2.0$ is necessary to reach the absolute
height of the data, as expected from the triple-differential results in the last section. Note
that the result of the correct approach and the standard (collinear) parton model coincide.
\subsection{E439 - $M$-spectrum}
\label{subsec:res_E439}

Experiment E439 measured dimuon production in pW collisions at $S \approx 750$ GeV$^2$.
The double differential cross section
\begin{equation}
	\frac{\diffd \sigma}{\diffd M \diffd x_F'}          \label{E439_double_xsec}								   
\end{equation}	
has been given with 
\begin{equation}
	x_F' = \frac{x_F}{1-\frac{M^2}{S}}
\end{equation}
at a fixed $x_F' = 0.1$.

As before we begin with Eqs.\ (\ref{master_naive}) and (\ref{master_correct}) and calculate the quantity 
Eq.\ (\ref{E439_double_xsec}) by integrating over $p_T^2$ and performing a simple transformation
from $x_F$ to $x_F'$:

\newpage
\begin{widetext}
\begin{align}
	\frac{\diffd \sigma}{\diffd M \diffd x_F'}
 	&=\int_0^{(p_T)^2_\textrm{max}} \diffd p_T^2
	      \frac{\diffd \sigma}{\diffd M \diffd x_F' \diffd p_T^2}
	=\int_0^{(p_T)^2_\textrm{max}} \diffd p_T^2\ 
	 2 M \left(1-\frac{M^2}{S}\right) \frac{\diffd \sigma}{\diffd M^2 \diffd x_F \diffd p_T^2} 
           \left(M, x_F=x_F'\left(1-\frac{M^2}{S}\right) \right) \ .
\end{align}
\end{widetext}

\begin{figure}[H]
	\centering
	\includegraphics[angle=-90,keepaspectratio,width=0.45\textwidth]
                        {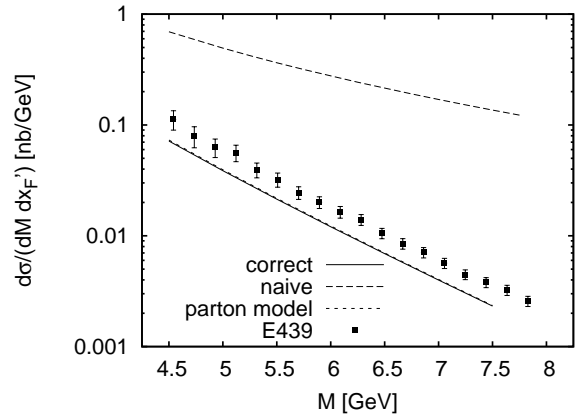}
	\caption{$M$-spectrum obtained from naive and correct approach with $D=0.5$
	         and from the standard parton model. Note that the results from the latter
		 two approaches are basically on top of each other.
		 Data are from E439 with $x_F' = 0.1$.}
	\label{fig:E439_double}	 
\end{figure}

We plot the results in Fig. \ref{fig:E439_double}, the solid line represents
the correct approach, the long dashed line the naive one. With the same parameter $D=0.5$ GeV
as for the E866 case we find the same discrepancy between the two approaches.
Again the correct approach reproduces the slope well and a factor $K\approx 1.6$ is 
required to fit the data, while the naive approach fails to describe the slope and
absolute size of the cross section. Once more the result of the correct approach
agrees well with the result of the standard parton model (short dashed).

Here we note the following: in \cite{Linnyk:2004mt,Linnyk:2005iw,Linnyk:2006mv} the
same data of experiment E439 are compared to calculations, however only to an approach
including both initial quark transverse momentum and quark mass distributions.
There it is found that the data can be described well without a K-factor. There is no
comparison of E439 data with a transverse momentum dependent calculation with onshell quarks
(i.e.\ what we call the naive approach) in \cite{Linnyk:2004mt,Linnyk:2005iw,Linnyk:2006mv}.
We acknowledge that the introduction of quark mass distributions lowers the cross section 
which somewhat compensates for the enhancement in the naive approach.       
Nonetheless we want to point out that even with additional smearing from the quark mass
distributions the naive approach will always lead to the wrong $M$-dependence of the
cross section. The reason is simply that the PDFs are probed in two areas: around the
standard collinear parton model $x$, cf.\ Eq.\ (\ref{spm_hadr_xsec}), and in a region very close
to $x=0$ where the PDFs behave very differently with $M$ and give much larger contributions
than for the physical $x$, since the PDFs diverge rapidly for $x \rightarrow 0$. Thus we conclude
that the agreement of the full calculation with the E439 data in 
\cite{Linnyk:2004mt,Linnyk:2005iw,Linnyk:2006mv} must be erroneous.

\section{Conclusions}
\label{sec:conc}

In this paper we reexamined a phenomenological model of Drell-Yan pair production
\cite{Linnyk:2004mt,Linnyk:2005iw,Linnyk:2006mv}. This model tried to describe the DY process
in a parton model scheme which takes into account the full transverse parton kinematics
in the hard subprocess. The aim was to reproduce the transverse momentum spectra of the
DY pairs and in addition to reproduce the absolute size of the cross sections by introducing
mass distributions of the partons.

We have shown that already in the first step of introducing full transverse kinematics important
constraints were not considered. Unphysical solutions emerging in a (too) naive parton model
contaminate the results. We have derived these constraints in the usual collinear approach
and then made the connection to the more general case of full kinematics. It turned out that 
unawareness of these constraints can lead to a drastically different result of the calculations:
while
in \cite{Linnyk:2004mt,Linnyk:2005iw,Linnyk:2006mv} the inclusion of the transverse kinematics
in the subprocess leads to an overshoot of the cross section, our corrected approach shows
no such behavior and instead nicely reproduces the standard parton model prediction for the
invariant mass spectra and the low transverse momentum spectrum, however only up to a 
K factor. Additionally we find that the naive approach taken in 
\cite{Linnyk:2004mt,Linnyk:2005iw,Linnyk:2006mv} produces a wrong $M$-dependence of the
cross section. This is a crucial point since already the standard parton model reproduces
the right slope of the $M$-spectrum. Therefore we conclude that the findings in 
\cite{Linnyk:2004mt,Linnyk:2005iw,Linnyk:2006mv}
that allow for a K factor free description of DY pair production are unwarranted.

\section*{Acknowledgements}

The authors are grateful to Kai Gallmeister for very helpful discussions.
F.E.\ was supported by DFG and S.L.\ was supported by GSI. This publication
represents a component of my doctoral (Dr. rer. nat.) thesis in the Faculty
of Physics at the Justus-Liebig-University Giessen, Germany.

\begin{appendix}

\section{Naive calculation of the hadronic cross section with full kinematics}

\label{appendix:naive}

Rewriting the $\delta$-functions of Eq. (\ref{naive_full_integral}) in terms of the
integration variables yields
\begin{widetext}
\begin{align}
	&\delta\left(M^2 - (p_1+p_2)^2\right)\,
	 \delta\left(x_F - \frac{(p_1 + p_2)_z}{(q_z)_\textrm{max}}\right)\,
         \delta\left(p_T^2 - ({\vec p}_{1_\perp} + {\vec p}_{2_\perp})^2 \right) \nonumber \\
	=\,
	&\delta\left(M^2 - p_1^+ p_2^- - p_1^- p_2^+ 
		     + 2 {\vec p}_{1_\perp} \cdot {\vec p}_{2_\perp}\right)\,
	 \delta\left(x_F - \frac{p_1^+ - p_1^- + p_2^+ - p_2^-}{2(q_z)_\textrm{max}}\right)\,
	 \delta\left(p_T^2 - ({\vec p}_{1_\perp})^2 - ({\vec p}_{2_\perp})^2 -  
		                         2 {\vec p}_{1_\perp} \cdot {\vec p}_{2_\perp}\right) 
							\nonumber \\
	=\,
	&\delta\left(M^2 - \frac{({\vec p}_{1_\perp})^2}{p_1^-}\frac{({\vec p}_{2_\perp})^2}{p_2^+}
	     - p_1^- p_2^+ + p_T^2 - ({\vec p}_{1_\perp})^2 - ({\vec p}_{2_\perp})^2\right)\,
	\delta\left(x_F - \frac{\frac{({\vec p}_{1_\perp})^2}{p_1^-} - p_1^- 
			 + p_2^+ - \frac{({\vec p}_{2_\perp})^2}{p_2^+}}{2(q_z)_\textrm{max}}
	                                                      \right) \nonumber \\
	&\times\delta\left(p_T^2 - ({\vec p}_{1_\perp})^2 - ({\vec p}_{2_\perp})^2 -  
		                         2 {\vec p}_{1_\perp} \cdot {\vec p}_{2_\perp}\right) 
					\nonumber \\
	=\,
	&\delta\left(M^2 - \frac{({\vec p}_{1_\perp})^2}{x_1 \sqrt{S}}
			   \frac{({\vec p}_{2_\perp})^2}{x_2 \sqrt{S}}
		         - x_1 \sqrt{S} x_2 \sqrt{S} + p_T^2 
			 - ({\vec p}_{1_\perp})^2 - ({\vec p}_{2_\perp})^2\right)\,
	\delta\left(x_F - \frac{\frac{({\vec p}_{1_\perp})^2}{x_1 \sqrt{S}} - x_1 \sqrt{S} 
			        + x_2 \sqrt{S} - \frac{({\vec p}_{2_\perp})^2}{x_2 \sqrt{S}}}
			       {2(q_z)_\textrm{max}}
	                                                      \right)\, \nonumber \\
	&\times \delta\left(p_T^2 - ({\vec p}_{1_\perp})^2 - ({\vec p}_{2_\perp})^2 -  
	                         2 {\vec p}_{1_\perp} \cdot {\vec p}_{2_\perp}\right)\ .
		\label{naive_delta}
\end{align}
Note here that if one puts all transverse momenta in Eq.
(\ref{naive_delta}) to zero the collinear relations (\ref{invmass_cond},\ref{feynx_cond})
are recovered. Note also that the unphysical parts of Eqs. (\ref{invmass_cond}) and (\ref{feynx_cond})
are included here since 
\begin{align}
	\tilde x_1 &= \frac{p_1^+}{\sqrt{S}} \rightarrow 
	              \frac{({\vec p}_{1_\perp})^2}{x_1 \sqrt{S}} \ , \label{full_naive_x1} \\
	\tilde x_2 &= \frac{p_2^-}{\sqrt{S}} \rightarrow 
	              \frac{({\vec p}_{2_\perp})^2}{x_2 \sqrt{S}} \ . \label{full_naive_x2}
\end{align}		      
Now using the first two $\delta$-functions of Eq. (\ref{naive_delta}) we can 
obtain solutions for the squared transverse momenta:
\begin{align}
	&\delta\left(M^2 - \frac{({\vec p}_{1_\perp})^2}{x_1 \sqrt{S}}
			   \frac{({\vec p}_{2_\perp})^2}{x_2 \sqrt{S}}
		         - x_1 \sqrt{S} x_2 \sqrt{S} + p_T^2 
			 - ({\vec p}_{1_\perp})^2 - ({\vec p}_{2_\perp})^2\right)\,
	\delta\left(x_F - \frac{\frac{({\vec p}_{1_\perp})^2}{x_1 \sqrt{S}} - x_1 \sqrt{S} 
			        + x_2 \sqrt{S} - \frac{({\vec p}_{2_\perp})^2}{x_2 \sqrt{S}}}
			       {2(q_z)_\textrm{max}} \right) \nonumber \\
	=\,
	& (q_z)_\textrm{max}\, \frac{x_1 x_2 S}{E}\, 
	  \delta\left(({\vec p}_{1_\perp})^2 - \left(E-x_2\sqrt{S} + (q_z)_\textrm{max}x_F\right)
			  						x_1 \sqrt{S}\right)\,
	  \delta\left(({\vec p}_{2_\perp})^2 - \left(E-x_1\sqrt{S} - (q_z)_\textrm{max}x_F\right)
			  						x_2 \sqrt{S} \right)
	  \ ,
	  \label{p1psq_p2psq}
\end{align}	
with the energy of the virtual photon
\begin{equation}
	E = \sqrt{M^2 + p_T^2 + (q_z)^2_\textrm{max}x_F^2}\ .
\end{equation}	
We note that exactly at this point the physical and the unphysical solutions for the momentum
fractions $x_i$ have been mixed up by rewriting the $\delta$-functions, since the unphysical
solutions of Eqs. (\ref{full_naive_x1}) and (\ref{full_naive_x2}) have entered.

Transforming the transverse momentum integrals
\begin{equation}
	\int \textrm{d}{\vec p}_{1_\perp} \int \textrm{d}{\vec p}_{2_\perp}
	= \int_{-\pi}^{\pi} \diffd {\phi}_{1_\perp} \int_0^{2\pi} \diffd {\phi}_{2_\perp}
	  \int_0^\infty \frac{1}{2} \textrm{d}({\vec p}_{1_\perp})^2 \int_0^\infty 
	  \frac{1}{2}\textrm{d}({\vec p}_{2_\perp})^2
\end{equation}	
we can rewrite the entire expression (\ref{naive_full_integral}) in the following form:
\begin{align}
	\frac{\textrm{d} \sigma_\textrm{naive}}{\diffd M^2\ \diffd x_F \diffd p_T^2} = 
	                  &\int_0^1 \textrm{d}x_1 \int_0^1 \textrm{d}x_2
	                   \int_{-\pi}^{\pi} \diffd {\phi}_{1_\perp} 
			   \int_0^{2\pi} \diffd {\phi}_{2_\perp}
	  		   \int_0^\infty \frac{1}{2} \textrm{d}({\vec p}_{1_\perp})^2 
			   \int_0^\infty \frac{1}{2}\textrm{d}({\vec p}_{2_\perp})^2
			   \nonumber \\ 
			   \times\,
			   &\frac{(q_z)_\textrm{max}{x_1 x_2 S}}{E}\,
			   \frac{F(x_1,{\vec p}_{1_\perp},x_2,{\vec p}_{2_\perp},q^2)}
                                {\sqrt{4({\vec p}_{1_\perp})^2({\vec p}_{2_\perp})^2
			   -\left(p_T^2 - ({\vec p}_{1_\perp})^2 -({\vec p}_{2_\perp})^2\right)^2}}
			   \nonumber   \\ 
			   \times\,
			   &\delta\left(({\vec p}_{1_\perp})^2 - 
			   \left(E-x_2\sqrt{S} + (q_z)_\textrm{max}x_F\right) x_1 \sqrt{S}\right)\,
	                    \delta\left(({\vec p}_{2_\perp})^2 - 
   			     \left(E-x_1\sqrt{S} - (q_z)_\textrm{max}x_F\right) x_2 \sqrt{S}\right)
			   \nonumber \\
			   \times\,	   
			   &\delta\left({\phi}_{1_\perp} - 
               \arccos\left(\frac{p_T^2 - \left({\vec p}_{1_\perp}\right)^2 
		                        - \left({\vec p}_{2_\perp}\right)^2}
	                    {\sqrt{4\left({\vec p}_{1_\perp}\right)^2
			            \left({\vec p}_{2_\perp}\right)^2}}\right)\right)
			\ .
\end{align}
Now all four integrations concerning the partons' transverse momenta can be carried out, leaving
a two-dimensional integral which must be calculated numerically:
\begin{equation}
		\frac{\textrm{d} \sigma_\textrm{naive}}{\diffd M^2\ \diffd x_F \diffd p_T^2} =
			   \int \textrm{d}x_1 \int \textrm{d}x_2
			   \frac{\pi\, (q_z)_\textrm{max}{x_1 x_2 S}}{E}\,
			   \frac{F(x_1,\hat{ \vec p}_{1_\perp},x_2,\hat{ \vec p}_{2_\perp},q^2)}
                                {\sqrt{4\left(\hat{ \vec p}_{1_\perp}\right)^2
					\left(\hat{ \vec p}_{2_\perp}\right)^2
			   -\left(p_T^2 - \left(\hat{ \vec p}_{1_\perp}\right)^2 
					- \left(\hat{ \vec p}_{2_\perp}\right)^2\right)^2}}
			   \label{master_naive}
			   \ .
\end{equation}
$\left(\hat{ \vec p}_{1_\perp}\right)^2$ and $\left(\hat{ \vec p}_{2_\perp}\right)^2$ 
are given by the $\delta$-functions in
Eq. (\ref{p1psq_p2psq}) and the integration boundaries of $x_1$ and $x_2$ have to be chosen
such that the requirements 
\begin{align}
	(\hat{ \vec p}_{1_\perp})^2 &> 0 \ ,\\
	(\hat{ \vec p}_{2_\perp})^2 &> 0 \ , \\
	4(\hat{\vec p}_{1_\perp})^2(\hat{ \vec p}_{2_\perp})^2 -\left(p_T^2 - (\hat{ \vec p}_{1_\perp})^2
-(\hat{ \vec p}_{2_\perp})^2\right)^2 &> 0 
\end{align}       
are fulfilled. One finds
\begin{equation}
	0 < x_1 < \frac{E - x_F(q_z)_\textrm{max}}{\sqrt{S}}
\end{equation}	
and
\begin{align}
\frac{x_1 \sqrt{S} (M^2-p_T^2) + p_T^2\, (E-x_F(q_z)_\textrm{max})
      - 2 M p_T \sqrt{x_1 \sqrt{S}  (E-x_F(q_z)_\textrm{max}-x_1\sqrt{S})}}
     {\sqrt{S} (E-x_F(q_z)_\textrm{max})^2} \nonumber 
\end{align}
\begin{align}
< x_2 < \nonumber
\end{align}
\begin{align}
\frac{x_1 \sqrt{S} (M^2-p_T^2) + p_T^2\, (E-x_F(q_z)_\textrm{max})
      + 2 M p_T \sqrt{x_1 \sqrt{S}  (E-x_F(q_z)_\textrm{max}-x_1\sqrt{S})}}
     {\sqrt{S} (E-x_F(q_z)_\textrm{max})^2} \ .
\end{align}

\end{widetext}

\end{appendix}

\bibliography{literature}
\bibliographystyle{apsrev4-1}

\end{document}